# CREATION OF OXIDE COATING ON AL 1050 ALLOY.

Alexander Sobolev[1], Alexey Kossenko[1], Michael Zinigrad[1], Konstantin Borodianskiy[1,*]

[1] Zimin Advanced Materials Laboratory, Department of Chemical Engineering, Biotechnology and Materials, Ariel University, Ariel 40700, Israel.

\* Correspondence: konstantinb@ariel.ac.il; Tel.: +972-3-9143085

**Abstract:** Plasma electrolytic oxidation (PEO) is a surface processing for oxide coatings formation with advanced properties. Mostly, PEO is performed in an aqueous solution electrolyte which limits the dimension of treated parts due to the system heating up. Therefore, treatment of large surfaces cannot be applied in aqueous electrolyte. In current work an alternative approach of PEO treatment applied in aluminum 1050 alloy in nitrate molten salt was investigated. The structure, phase and chemical compositions, and micro-hardness were examined using XRD, SEM, EDS and micro-hardness tests. The obtained results showed that formed coating contains from two phases of α- $Al_2O_3$ and γ- $Al_2O_3$. It was found that formed coating is free of any contaminants originated from the electrolyte and has no damages, which are usually present in coatings formed in PEO treatment in aqueous solution electrolyte.

## 1. Introduction

In last twenty years plasma electrolytic oxidation (PEO) became one of the attractive coating approaches [1-5]. It is usually applied in so called valve metals as aluminum, magnesium or titanium to obtain surface required properties as high electrical insulation, corrosion and wear resistance, as well as excellent performance [6-8]. It is usually applied in medical, oil and gas, automotive, and aerospace industries.

The basic principle of the PEO is a high voltage application between the specimen subjected to the treatment and the electrode, while the micro-arc discharge migration points appear on the surface [9-11].

The majority of research works describe PEO process in different aqueous solutions [12]. Some of them show the process of *in situ* doping with cerium or phosphate [13], tungsten [14] or even nanoplatelets [15]. Other works describe nanoformation on the coating surface which affects its final properties [16-17]. The main disadvantages of the PEO treatment in aqueous solutions are a relatively low coating rate, a formation of a thick high porosity layer, a necessity of electrolyte cooling, and a presence of some undesirable compounds in the coating [18]. These issues can be solved by performing PEO in molten salt [19-20].

In presented work a ceramic coating formation achieved by PEO treatment in nitrate molten salt of the eutectic composition was studied. The obtained morphology, chemical and phase composition, and performance were discussed in the work. Moreover, different stages of the treatment and their comparison to the process in aqueous solution discussed as well.

## 2. Materials and Methods

Al alloy 1050 specimens (chemical composition shown in Table 1) with a surface area of 0.16-0.17 dm$^2$ were grinded by using of abrasive papers grits #280, #400, #600, #1000, #2400 and #4000, and then subjected to ultrasonically cleaning in acetone. The surface roughness after polishing was Ra 3.

**Table 1.** Chemical composition of the Al alloy 1050.

| Chemical element, % mass | | | | | | | |
|---|---|---|---|---|---|---|---|
| Si | Mg | Fe | Cu | Mn | Zn | Ti | Al |
| 0.25 | 0.05 | 0.44 | 0.05 | 0.05 | 0.07 | 0.05 | Base |

PEO treatment was performed at 280˚C in the electrolyte with the eutectic composition of $NaNO_3 – KNO_3$. The electrolyte was held in a Ni crucible (99.95% Ni) which served as a counter electrode. The anodic current density was 70mA/cm$^2$, voltage was limited by the galvanostatic mode. The applied power supply has the following parameters: $I_{max}$ = 5A, $U_{max}$ =900V, current and voltage were pulsed with a square-wave sweep at a frequency of 50Hz ($t_a$ = $t_k$ = 0.01s) by a pulse generator. Duration of PEO treatment was 10min with the coating rate of 1μm/min.

The current and the voltage wave profiles and trend plots, as well as the power consumed during the process, were measured using Fluke Scope Meter 199C (200 MHz, 2.5 GS s$^{-1}$), which was located into the electrical circuit between the power supply and the working container.

Cross-sections were prepared by a standard method for metallography. Surface and cross-section morphologies of the obtained PEO coatings were examined by TESCAN MAIA3 SEM equipped with an energy dispersive X-ray spectroscopy (EDS) system. The phase composition of the coatings were determined by a PANalytical Empyrean diffractometer (in Cu Kα radiation) in grazing incidence mode using a scan with a grazing angle of 3°, a step size of 0.03°, and at a 2Θ range from 10° to 90°.

A BuehlerMicromet 2100 microhardness tester was used to evaluate microhardness on a cross section of the obtained oxide layer.

## 3. Results

Figure 1 shows the voltage and the current behavior as a function of treatment time.

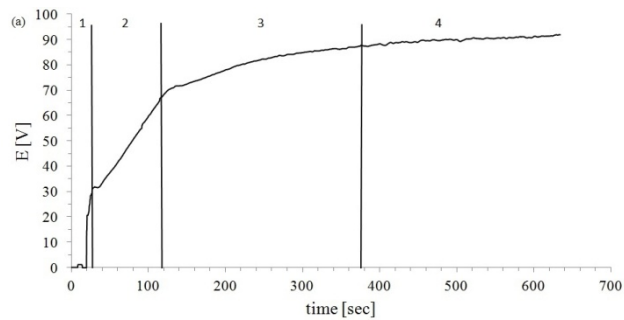

**Figure 1.** Electric parameters behavior plot voltage as a function of treatment time.

Figure 1 demonstrates different stages of oxide layer formation that corresponds to the different stages on the voltage – time plot: on region 1 no discharges were revealed; region 2 corresponds to the microarc discharges start to appear; region 3-4 corresponds to the microarc discharges number increases.

Surface morphology and chemical composition (EDS) of the sample were investigated as well and the obtained images by SEM are presented in Figure 2.

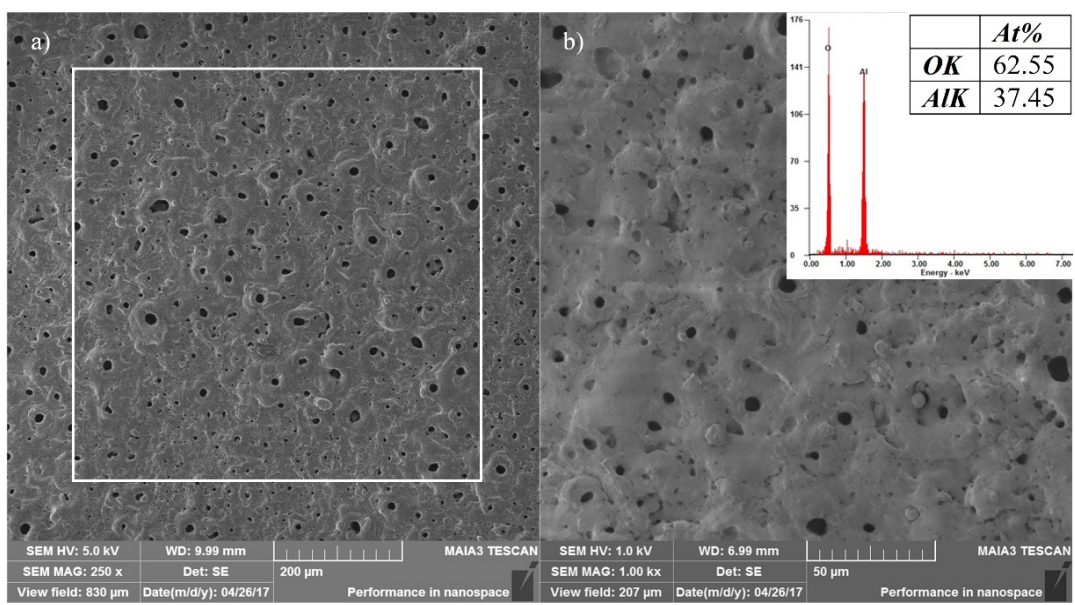

**Figure 2.** SEM image of surface morphology of the alloy Al 1050 treated by PEO with magnifications: (a) ×250, (b) ×1000, and the chemical composition obtained by EDS (b).

The presented micrographs show that the surface morphology of the treated alloy contains round shaped pores with a diameter at the range of 0.5 - 5µm. That's correlated with the treatment obtained in aqueous solution and compared later in discussions of the current work.

EDS was performed for qualitative analysis; however quantification was performed as well and consequently followed by XRD investigation to evaluate coating phase composition. According to EDS investigations, atomic composition of aluminum and oxygen are 37.45% and 62.55% respectively.

The obtained semi-quantitative calculations by XRD show phase composition distribution of the coating. Results in Figure 3 indicate the presence of 55% of α-$Al_2O_3$ and 45% of γ-$Al_2O_3$.

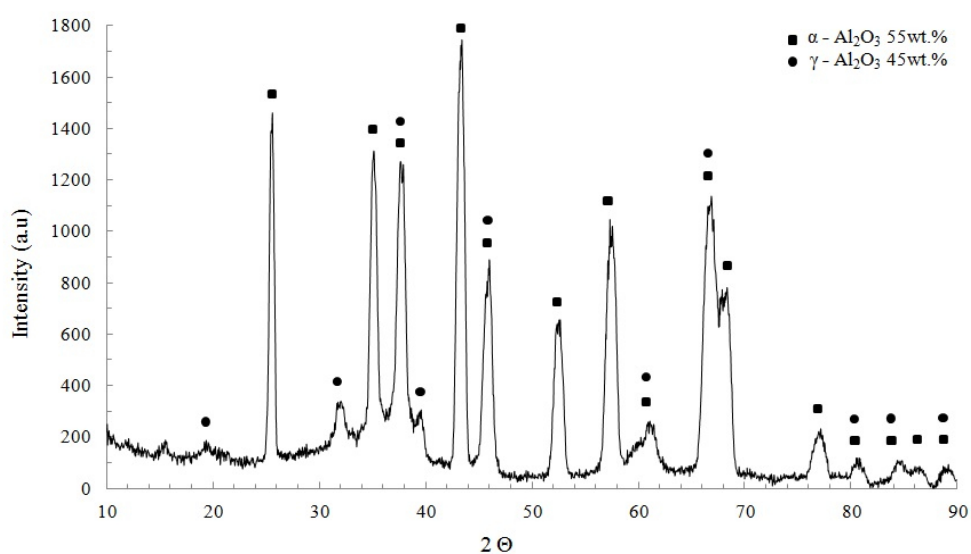

**Figure 3.** XRD pattern of Al 1050 alloy surface after PEO treatment.

Micro-hardness measurements were performed on the oxide coating and the obtained results are illustrated in Table 2. Different values of the micro-hardness obtained due to the different layers composition depends on the depth.

**Table 2.** Micro-hardness test results of different alloy coating depth related to Figure 4: Base metal – point 4, inner layer – point 3, outer layer – point 2.

| Base aluminum [$HV_{10}$] | Inner layer [$HV_{10}$] | Outer layer [$HV_{10}$] |
|---|---|---|
| 40 ± 2.2 | 768 ± 2.1 | 1048 ± 2.2 |

Figure 4 demonstrates cross section micrograph of the obtained coating after PEO treatment and its EDS line scan elemental analysis.

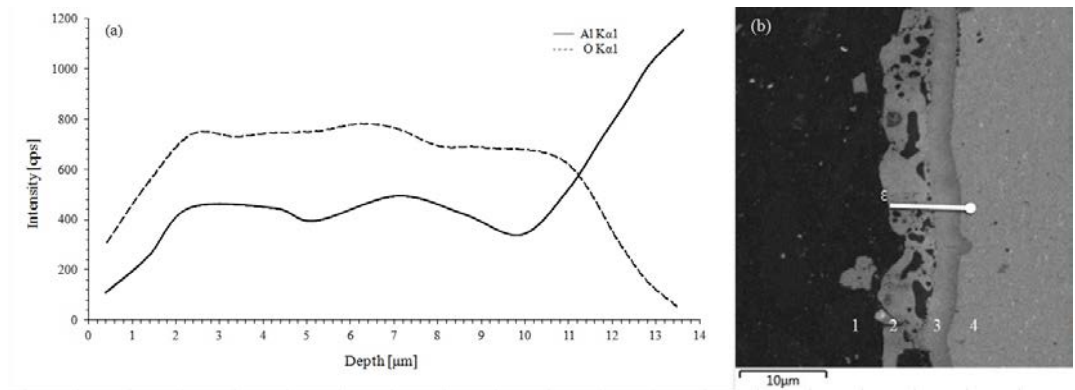

**Figure 4.** EDS line scan of the aluminum 1050 alloy cross section after PEO treatment and its microphotograph: 1 – resin; 2 – outer layer; 3 – inner layer; 4 – aluminum base.

SEM image show that the thickness of the porous outer layer is about 4µm and the thickness of the inner dense layer is about 3.5µm. From the EDS results, we found that oxygen (O) and aluminum (Al) concentration fit the oxide layer of $Al_2O_3$ composition that is also indicated by XRD analysis.

## 4. Discussion

Evaluation of morphological changes showed that formed alloy subjected to the PEO treatment has a round shaped porosity with the diameter at the range of 0.5 - 5µm. Micrographs shown illustrated porous structure of the outer layer while the inner has no porosity at all. SEM images showed the morphology of the obtained coating and they are correlated with the morphology usually obtained in PEO process performed in aqueous solution electrolyte [21].

No cracks were found in the formed coating. We suggest that the low cooling rate is responsible for this. However, coating produced in aqueous solution usually formed from the oxide layer with cracks; that attributed to the high cooling rate of the coating formation [22].

EDS analysis, line scan and XRD investigations evaluated the coating which is free of any contamination in contrast to the coating obtained in aqueous solution where contamination is originated from the electrolyte [23]. Typically, electrolytes for aluminum alloys PEO treatment made of alkali metal hydroxides and sodium silicate. Therefore, a formed crystalline aluminum oxide coating contains of Si in a form of oxide.

XRD results demonstrate presence of α-Al$_2$O$_3$ and γ-Al$_2$O$_3$ phases in the obtained coating. The micro-hardness values of α-Al$_2$O$_3$ are higher than of the γ-Al$_2$O$_3$, therefore the outer layer consists of α-Al$_2$O$_3$ while the inner of γ-Al$_2$O$_3$. No other phases were detected; that's additionally approves the statement done above. These results are well correlated with the micro-hardness values mentioned in Table 2 and in presented images obtained by electron microscopy.

## 5. Conclusions

In current work a new approach of oxide coating formation in PEO treatment in nitrate molten salt of the eutectic composition was investigated. The following results and conclusions were done:

1. Aluminum 1050 alloy subjected to PEO treatment has the oxide coating contained from two sub-layers α- Al$_2$O$_3$ and γ- Al$_2$O$_3$.
2. The formed coating has an oxide composition and it's free of any contaminants originated from the electrolyte. Additionally, no through porosity and cracks were found.
3. The proposed process can be applied to the high surface area coating in contrast to the treatment performed in aqueous solutions.

**Acknowledgments:** Authors would like to thank Ms. Natalia Litvak for her help in electron microscopy investigation.